\documentclass[12pt]{article}
\usepackage{amssymb,amsmath,epsfig}
\usepackage{graphicx}
\allowdisplaybreaks

\begin{document}
\title{\bf Viable and Stable Compact Stellar Structures in $f(\mathcal{Q},\mathcal{T})$ Theory}
\author{M. Zeeshan Gul \thanks{mzeeshangul.math@gmail.com},
M. Sharif \thanks {msharif.math@pu.edu.pk}~
and Adeeba Arooj \thanks{aarooj933@gmail.com}\\
Department of Mathematics and Statistics, The University of Lahore,\\
1-KM Defence Road Lahore-54000, Pakistan.}
\date{}
\maketitle

\begin{abstract}
The main objective of this paper is to investigate the impact of
$f(\mathcal{Q},\mathcal{T})$ gravity on the geometry of anisotropic
compact stellar objects, where $\mathcal{Q}$ is non-metricity and
$\mathcal{T}$ is the trace of the energy-momentum tensor. In this
perspective, we use the physically viable non-singular solutions to
examine the configuration of static spherically symmetric
structures. We consider a specific model of this theory to examine
various physical quantities in the interior of the proposed compact
stars. These quantities include fluid parameters, anisotropy, energy
constraints, equation of state parameters, mass, compactness and
redshift. The Tolman-Oppenheimer-Volkoff equation is used to examine
the equilibrium state of stellar models, while the stability of the
proposed compact stars is investigated through sound speed and
adiabatic index methods. It is found that the proposed compact stars
are viable and stable in the context of this theory.
\end{abstract}

\section{Introduction}

Einstein's general theory of relativity (GR) is a fundamental theory
that provides a new understanding of gravity and the nature of
spacetime. It remains one of the pillars of modern physics and has
been extensively tested through various observations and
experiments. As GR is based on geometric structures in Riemann's
metric space, an alternative approach to generalize GR is to use
more general geometrical structures that could explain the
gravitational field and describe the behavior of matter at large
cosmic scales. In this perspective, Weyl \cite{2} attempted to
develop a more general geometry than Riemannian space. This approach
has the objective to unify these fundamental forces under a single
geometric framework. In Riemannian geometry, an important concept is
the Levi-Civita connection, which compares vectors according to
their length. During parallel transport, Weyl introduced a
connection that does not contain information about the length of
vectors. To address the lack of information about vector length, he
introduced an additional connection known as the \emph{length
connection}. This connection was not concerned with the direction of
vector transport but rather with fixing or \emph{gauging} the
conformal factor.

Weyl's theory posits that the covariant divergence of the metric
tensor is non-zero, leading to the concept of non-metricity. He
identified length connection with electromagnetic potential for
physical applications. A theoretical gauge \cite{3} refers to a
mathematical framework that describes fundamental forces and fields
in physics. Non-Riemannian geometries can include concepts like
torsion and nonmetricity. The nonmetricity scalar is a mathematical
quantity that arises in theories involving non-Riemannian
geometries. In some contexts, it has been proposed as a way to
determine the cosmic expansion \cite{4}. While Einstein's
formulation of GR focuses on curvature, alternative theories
consider torsion and nonmetricity as additional geometrical
properties of spacetime. Teleparllel gravity is an alternative
theory to GR in which torsion $(T)$ represents the gravitational
interaction. In contrast to the Levi-Civita connection, the
teleparallel equivalent of GR exhibits nonmetricity and zero
curvature. To characterize GR for torsion curvature \cite{5} and
nonmetricity \cite{6}, the integral action of GR is expressed as
$\int \sqrt{-g}T$ and $\int \sqrt{-g}\mathcal{Q}$, respectively.

Yixin et al \cite{20} generalized $f(\mathcal{Q})$ theory of gravity
by incorporating the  trace of the energy-momentum tensor in the
functional action, known as $f(\mathcal{Q},\mathcal{T})$ gravity.
This theory establishes a specific coupling between trace of the
energy-momentum tensor and nonmetricity. The motivation behind
studying this theory includes exploring its theoretical
implications, its compatibility with observational data, and its
relevance in cosmological contexts. Arora et al \cite{22}
investigated whether this gravity can account for the late-time
acceleration of the universe without introducing additional forms of
dark energy. Bhattacharjee et al \cite{23} studied the phenomenon of
baryogenesis (generation of matter-antimatter asymmetry) in this
framework. This theory is reported to change the nature of tidal
forces and the equation of motion in the Newtonian limit, suggesting
deviations from classical predictions \cite{21}. Researchers aim to
compare predictions from $f(\mathcal{Q}, \mathcal{T})$ gravity,
particularly those related to tidal force changes, with observable
evidence from various astrophysical phenomena.

The formation and evolution of galaxies are complex processes that
involve the interplay of various astrophysical phenomena. Stars are
crucial components of galaxies and maintain a state of equilibrium
when the inner gravitational force is balanced by the outward
pressure exerted by the nuclear fusion reactions occurring in their
cores. Once the nuclear fuel is consumed, insufficient pressure
prevents the star from collapsing. Consequently, new dense stars are
formed, known as compact stars (CSs). The configuration of dense
objects inspired several researchers to analyze their different
evolutionary stages and interior attributes in the background of
astrophysics. In this regard, the exact composition and internal
structure of neutron stars have been the subject of extensive
research in gravitational physics.

Baade and Zwicky \cite{24} argued that CSs are formed because of
supernova and their existence has been proved by pulsars \cite{25}.
Pulsars are highly magnetized rotating neutron stars which emit
electromagnetic radiation beams. These beams are observed as regular
pulses of radiation as the neutron star rotates, hence the name
``pulsar." Studying neutron stars and pulsars allows scientists to
explore various aspects of these intriguing objects. Neutron stars
also provide valuable insights into fundamental physics such as the
behavior of matter under extreme densities and the effects of strong
gravitational fields. Neutron stars have attracted considerable
attention due to their fascinating properties and structures. Mak
and Harko \cite{26} investigated the viability of pulsars through
energy bounds and examined their stable state through sound speed.
Rahaman et al \cite{27} used the EoS parameter to analyze the viable
features of CSs.

The viable characteristics of CSs yield fascinating outcomes in the
framework of alternative theories of gravity. Arapoglu et al
\cite{28} used the perturbation technique to examine the geometry of
CSs in $f(\mathcal{R})$ gravity. Astashenok et al \cite{29}
discussed the structure of pulsars by analyzing the profile of
matter contents in the same theory. Das et al \cite{30} examined the
impact of effective matter variables on the geometry of anisotropic
relativistic sphere in $f(\mathcal{R}, \mathcal{T})$ theory. Deb et
al \cite{31} analyzed the geometry of spherically symmetric
isotropic strange stars to study the viability of the considered
stellar models in the same theory. Biswas et al \cite{32} discussed
the strange quark stars admitting the Krori-Barua solution in this
theory. Bhar et al \cite{33} used the Tolman-Kuchowicz solution to
investigate the viable characteristics of 4U 1538-52 CS in Einstein
Gauss-Bonnet gravity. Sharif and Ramzan \cite{34} studied the
behavior of various physical quantities and stability of distinct
CSs in $f(\mathcal{G})$ theory. Dey et al \cite{36} considered
Finch-Skea ansatz to study the viable anisotropic stellar models in
$f(\mathcal{R}, \mathcal{T})$ theory. Sharif and Gul \cite{37}
studied the physical attributes of CSs through Noether symmetry
approach in $f(\mathcal{R}, \mathcal{T}^{2})$ theory.

The above literature motivates us to investigate the viable
characteristics of anisotropic CSs in $f(\mathcal{Q},\mathcal{T})$
gravity. We use the following format in the paper. The basic
formulation of $f(\mathcal{Q},\mathcal{T})$ gravity is given in
section \textbf{2}. In section \textbf{3}, we consider a specific
model of this theory to formulate the explicit expression of energy
density and pressure components. Also, we evaluate unknown
parameters through the matching conditions. Section \textbf{3}
determines physical features of the considered CSs through different
physical quantities. The equilibrium state and stability of the
considered CSs are analyzed in section \textbf{4}. We compile our
outcomes in section \textbf{5}.

\section{Basic Formalism of $f(\mathcal{Q},\mathcal{T})$ Theory}

This section presents the fundamental framework of the modified
$f(\mathcal{Q},\mathcal{T})$ theory and derived the field equations
by variational principle. Weyl \cite{2} introduced a generalization
of Riemannian geometry as a mathematical framework for describing
gravitation in GR. In Riemannian geometry, parallel transport around
a closed path preserves a vector's direction and length. Weyl
proposed a modification where a vector would change its direction
and length during parallel transport around a closed path. This
modification involves a new vector field $(k^{\alpha})$ which
characterizes the geometric properties of Weyl geometry. The
fundamental fields in Weyl's space are the new vector field and
metric tensor. The metric tensor determines the local structure of
spacetime, defining distances and angles, while the vector field is
introduced to account for the change in length during parallel
transport. According to Weyl's theory, vector field has the same
mathematical properties as electromagnetic potentials in physics,
which indicates a strong connection between gravitational and
electromagnetic forces. Both forces are long-range forces and Weyl's
proposal raises the possibility of a common geometric origin for
these forces \cite{38}.

In a Weyl geometry, if a vector of length $y$ is transported with an
infinitesimal path $\delta x^{\alpha}$ then its length changes as
$\delta y= yk_{\alpha} \delta x^ {\alpha}$ \cite{38}. This indicates
that the variation in the vector's length is proportional to the
original length, the connection coefficient and the displacement
along the path. The variation in the vector's length after it is
transported in parallel around a tiny closed loop with area $\delta
s^{\alpha\beta}$ in the Weyl space is given as $\delta y = y
\Psi_{\alpha\beta}\delta s^{\alpha\beta}$, where
\begin{equation}\label{1}
\Psi_{\alpha\beta}=\nabla_{\beta}k_{\alpha}-\nabla_{\alpha}k_{\beta}.
\end{equation}
This states that the variation in the vector's length is
proportional to the original length, the curvature of the Weyl
connection and the area enclosed by the loop. A local scaling length
of the form $\tilde{y}=\varepsilon(x)y$ changes the field equation
$\tilde{k_{\alpha}}$ to $\tilde{k_{\alpha}}=k_{\alpha}+
(\ln\varepsilon),_{\alpha}$, whereas the elements of metric tensor
are modified by the conformal transformations
$\tilde{g}_{\alpha\beta} = \varepsilon^{2} g_{\alpha\beta}$ and
$\tilde{g}^{\alpha\beta} = \varepsilon^{-2} g^{\alpha\beta}$,
respectively \cite{39}. A semi-metric connection is another
important feature of the Weyl geometry, defined as
\begin{equation}\label{2}
{\bar{\Gamma}}^{\gamma}_{\alpha\beta}=\Gamma^{\gamma}_{\alpha\beta}
+g_{\alpha\beta}k^{\gamma}-\delta^{\gamma}_{\alpha}k_{\beta}-
\delta^{\gamma}_{\beta}k_{\alpha},
\end{equation}
where $\Gamma^{\gamma}_{\alpha\beta}$ denotes Christoffel symbol.
One can construct a gauge covariant derivative based on the
supposition that $ \bar{\Gamma}^{\gamma}_{\alpha\beta}$ is symmetric
\cite{39}. The Weyl curvature tensor using the covariant derivative
can be expressed as
\begin{equation}\label{3}
\bar{\mathcal{W}}_{\alpha\beta\gamma\eta}=\bar{\mathcal{W}}
_{(\alpha\beta)\gamma\eta}+\bar{\mathcal{W}}_{[\alpha\beta]
\gamma\eta},
\end{equation}
where
\begin{eqnarray}\nonumber
\bar{\mathcal{W}}_{[\alpha\beta]\gamma\eta}&=&\mathcal{W}
_{\alpha\beta\gamma\eta}+2\nabla_{\gamma}k_{[\alpha
g_{\beta}]\eta}+2\nabla_{\eta}k_{[\beta
g_{\alpha}]}\gamma+2k_{\gamma}k_{[\alpha
g_{\beta}]\eta}+2k_{\eta}k_{[\beta g_{\alpha}]\gamma}
\\\nonumber
&-&2k^{2}g_{\gamma[\alpha g_{\beta}]\eta},
\\\nonumber
\bar{\mathcal{W}}_{(\alpha\beta)\gamma\eta}&=&
\frac{1}{2}(\bar{\mathcal{W}} _{\alpha\beta\gamma\eta}
+\bar{\mathcal{W}}_{\beta\alpha\gamma\eta})
=g_{\alpha\beta}\Psi_{\gamma\eta}.
\end{eqnarray}
The Weyl curvature tensor after the first contraction yields
\begin{eqnarray}\label{4}
\bar{\mathcal{W}}^{\alpha}_{\beta}&=&\bar{\mathcal{W}}
^{\gamma\alpha}_{\gamma\beta}=\mathcal{W}^{\alpha}_{\beta}
+2k^{\alpha}k_{\beta}+3\nabla_{\beta}k^{\alpha}-\nabla_{\alpha}k^{\beta}
+g^{\alpha}_{\beta}(\nabla_{\gamma}k^{\gamma}-2k_{\gamma}k^{\gamma}).
\end{eqnarray}
Finally, we obtain Weyl scalar as
\begin{equation}\label{5}
\bar{\mathcal{W}}=\bar{\mathcal{W}}^{\gamma}_{\gamma}=
\mathcal{W}+6(\nabla_{\alpha}k^{\alpha}-k_{\alpha}k^{\alpha}).
\end{equation}

Weyl-Cartan (WC) spaces with torsion represent a more generalized
framework beyond Riemannian and Weyl geometry. This broader
geometric structure can be used to model theories of gravity that
include additional degrees of freedom beyond GR, allowing for
different scales and parallel transport behaviors. Such approaches
might be explored in the context of alternative theories of gravity
or in attempts to unify gravity with other fundamental forces. In a
WC spacetime, the length of a vector is defined by a symmetric
metric tensor and the law of parallel transport is determined by an
asymmetric connection as
$d\upsilon^{\alpha}=-\upsilon^{\gamma}\hat{\Gamma}^{\alpha}_{\gamma\beta}dx^{\beta}$
\cite{40}. The connection for the WC geometry is expressed as
\begin{equation}\label{6}
\hat{\Gamma}^{\gamma}_{\alpha\beta}={\Gamma}^{\gamma}_{\alpha\beta}
+\mathcal{C}^{\gamma}_{\alpha\beta}+\mathcal{L}^{\gamma}_{\alpha\beta},
\end{equation}
where $ \mathcal{C}^{\gamma}_{\alpha\beta}$ is the contortion tensor
and $ \mathcal{L}^{\gamma}_{\alpha\beta}$ is the disformation
tensor. The contorsion tensor from the torsion tensor can be
obtained as
\begin{equation}\label{7}
\mathcal{C}^{\gamma}_{\alpha\beta}=\hat{\Gamma}^{\gamma}_{[\alpha\beta]}
+g^{\gamma\eta}g_{\alpha\varsigma}
\hat{\Gamma}^{\varsigma}_{[\beta\eta]}+g^{\gamma\eta}
g_{\beta\varsigma} \hat{\Gamma}^{\varsigma}_{[\alpha\eta]}.
\end{equation}

The non-metricity yields the disformation tensor as
\begin{equation}\label{8}
\mathcal{L}^{\gamma}_{\alpha\beta}=\frac{1}{2}g^{\gamma\eta}
(\mathcal{Q}_{\beta\alpha\eta}
+\mathcal{Q}_{\alpha\beta\eta}-\mathcal{Q}_{\gamma\alpha\beta}),
\end{equation}
where
\begin{equation}\label{9}
\mathcal{Q}_{\gamma\alpha\beta}=\nabla_{\gamma} g_{\alpha\beta}
=-g_{\alpha\beta,\gamma}+g_{\beta\eta}\hat{\Gamma}^{\eta}_{\alpha\gamma}
+g_{\eta\alpha}\hat{\Gamma}^{\eta}_{\beta\gamma}.
\end{equation}
Here, $\hat{\Gamma}^{\gamma}_{\alpha\beta}$ is  WC connection. From
Eqs. (\ref{2}) and (\ref{6}), it is clear that the WC geometry with
zero torsion is a particular case of Weyl geometry, where the
non-metricity is defined as $\mathcal{Q}_{\gamma\alpha\beta} =
-2g_{\alpha\beta}k_{\gamma}$. Therefore, Eq.(\ref{6}) turns out to
be
\begin{equation}\label{10}
\hat{\Gamma}^{\gamma}_{\alpha\beta}={\Gamma}^{\gamma}_{\alpha\beta}
+g_{\alpha\beta}k^{\gamma}
-\delta^{\gamma}_{\alpha}k_{\beta}-\delta^{\gamma}_{\beta}k_{\alpha}
+\mathcal{C}^{\gamma}_{\alpha\beta},
\end{equation}
where
\begin{equation}\label{11}
\mathcal{C}^{\gamma}_{\alpha\beta}=T^{\gamma}
_{\alpha\beta}-g^{\gamma\eta}
g_{\varsigma\alpha}T^{\varsigma}_{\eta\beta}-g^{\gamma\eta}
g_{\varsigma\beta}T^{\varsigma}_{\eta\alpha},
\end{equation}
is the contortion and the WC torsion is expressed as
\begin{equation}\label{12}
T^{\gamma}_{\alpha\beta}=\frac{1}{2} (\hat{\Gamma}^{\gamma}
_{\alpha\beta}-\hat{\Gamma}^{\gamma}_{\beta\alpha}).
\end{equation}
The WC curvature tensor with the use of the connection is defined as
\begin{equation}\label{13}
\hat{\mathcal{W}}^{\gamma}_{\alpha\beta\eta} =\hat{\Gamma}^{\gamma}
_{\alpha\eta,\beta}
-\hat{\Gamma}^{\gamma}_{\alpha\beta,\eta}+\hat{\Gamma}
^{\varsigma}_{\alpha\eta}
\hat{\Gamma}^{\gamma}_{\varsigma\beta}-\hat{\Gamma}
^{\varsigma}_{\alpha\beta} \hat{\Gamma}^{\gamma}_{\varsigma\eta}.
\end{equation}
The WC scalar can be obtained by contracting the curvature tensor as
\begin{eqnarray}\nonumber
\hat{\mathcal{W}}&=&\hat{\mathcal{W}}^{\alpha\beta} _{\alpha\beta}
=\mathcal{W}+6\nabla_{\beta}k^{\beta}-4\nabla_{\beta}
T^{\beta}-6k_{\beta}k^{\beta}
+8k_{\beta}T^{\beta}+T^{\alpha\gamma\beta}T_{\alpha\gamma\beta}
\\\label{14}
&+&2T^{\alpha\gamma\beta}T_{\beta\gamma\alpha}-4T^{\beta}T_{\beta},
\end{eqnarray}
where$T_{\beta}=T^{\beta}_{\alpha\beta}$ and all covariant
derivatives are considered corresponding to metric.

The gravitational action can be reformulated by eliminating the
boundary terms in the Ricci scalar as \cite{41}
\begin{equation}\label{15}
\mathcal{S}=\frac{1}{2\kappa} \int
g^{\alpha\beta}(\Gamma^{\gamma}_{\eta\alpha}\Gamma^{\eta}_{\gamma\beta}
-\Gamma^{\gamma}_{\eta\gamma}\Gamma^{\eta}_{\alpha\beta})\sqrt{-g}
d^ {4}x.
\end{equation}
Based on the assumption that the connection is symmetric, we have
\begin{equation}\label{16}
\Gamma^{\gamma}_{\alpha\beta}=-\mathcal{L}^{\gamma}_{\alpha\beta}.
\end{equation}
Thus, the gravitational action becomes
\begin{equation}\label{17}
\mathcal{S}=-\frac{1}{2\kappa} \int
g^{\alpha\beta}(\mathcal{L}^{\gamma}_{\eta\alpha}
\mathcal{L}^{\eta}_{\gamma\beta} - \mathcal{L}^{\gamma}_{\eta\gamma}
\mathcal{L}^{\eta}_{\alpha\beta}) \sqrt{-g} d^ {4}x,
\end{equation}
where the non-mitricity scalar is defined as
\begin{equation}\label{18}
\mathcal{Q}\equiv-g^{\alpha\beta}(\mathcal{L}^{\gamma}_{\eta\alpha}
\mathcal{L}^{\eta}_{\gamma\beta}
-\mathcal{L}^{\gamma}_{\eta\gamma}\mathcal{L}^{\eta}_{\alpha\beta}),
\end{equation}
with
\begin{equation}\label{19}
\mathcal{L}^{\gamma}_{\eta\alpha}\equiv-\frac{1}{2}g^{\gamma\varsigma}
(\nabla_{\alpha}g_{\eta\varsigma}+\nabla_{\eta}g_{\varsigma\gamma}
-\nabla_{\varsigma}g_{\eta\alpha}).
\end{equation}
From Eq.(\ref{17}), one can obtain the gravitational action of
$f(\mathrm{Q})$ theory by replacing non-mitricity scalar with an
arbitrary function as
\begin{equation}\label{19a}
\mathcal{S}=\frac{1}{2\kappa}\int f(\mathcal{Q})\sqrt{-g}d^{4}x.
\end{equation}
This is the action of symmetric teleparallel theory, which is a
theoretical framework that provides an alternative geometric
description of gravity.

Now, we extend this gravitational Lagrangian by introducing the
trace of energy-momentum tensor in the functional action as
\begin{equation}\label{19b}
\mathcal{S}=\frac{1}{2\kappa}\int f(\mathcal{Q},\mathcal{T})
\sqrt{-g}d^{4}x.
\end{equation}
The modified Einstein-Hilbert action of $f(\mathcal{Q},\mathcal{T})$
gravity with geometric and matter part is defined as
\begin{equation}\label{20}
\mathcal{S}=\frac{1}{2\kappa}\int f(\mathcal{Q},\mathcal{T})
\sqrt{-g}d^{4}x+\int \mathcal{L}_{m}\sqrt{-g}d^{4}x,
\end{equation}
where $g$ is the determinant of the metric tensor, $\mathcal{L}_m$
represents the matter-lagrangian and $\kappa=1$ defines the coupling
constant. The trace of the non-metricity tensor is defined as
\begin{eqnarray}\label{21}
\mathcal{Q}_{\gamma}\equiv \mathcal{Q}^{~\alpha}_{\gamma~~\alpha},
\quad \tilde{\mathcal{Q}}_{\gamma}\equiv
\mathcal{Q}^{\alpha}_{\gamma\alpha}.
\end{eqnarray}
The superpotential of this model is expressed as
\begin{equation}\label{22}
\mathcal{P}^{\gamma}_{\alpha\beta}=-\frac{1}{2}\mathcal{L}
^{\gamma}_{\alpha\beta} +\frac{1}{4}(\mathcal{Q}^{\gamma}
-\tilde{\mathcal{Q}}^{\gamma})g_{\alpha\beta}- \frac{1}{4} \delta
^{\gamma} _{[\alpha \mathcal{Q}_{\beta}]}.
\end{equation}
The relation for $\mathcal{Q}$ is
\begin{equation}\label{23}
\mathcal{Q}=-\mathcal{Q}_{\gamma\alpha\beta}\mathcal{P}
^{\gamma\alpha\beta}=-\frac{1}{4}
(-\mathcal{Q}^{\gamma\beta\eta}
\mathcal{Q}_{\gamma\beta\eta}+2\mathcal{Q}^{\gamma\beta\eta}
\mathcal{Q}_{\eta\gamma\beta}
-2\mathcal{Q}^{\eta}\tilde{\mathcal{Q}}_{\eta}+\mathcal{Q}
^{\eta}\mathcal{Q}_{\eta}).
\end{equation}
The calculation of the above relation is shown in Appendix
\textbf{A}.

By varying Eq.(\ref{1}) with respect to the metric tensor, we obtain
\begin{eqnarray}\nonumber
\delta\mathcal{S}&=&\int \frac{1}{2} \delta [f(\mathcal{Q},T)
\sqrt{-g}] + \delta [\mathcal{L}_{M} \sqrt{-g}] d^ {4}x,
\\\nonumber
&=&\int \frac{1}{2}( \frac{-1}{2} f g_{\alpha\beta} \sqrt{-g} \delta
g^{\alpha\beta} + f_{\mathcal{Q}} \sqrt{-g} \delta \mathcal{Q} +
f_{\mathcal{T}} \sqrt{-g} \delta \mathcal{T})
\\\label{24}
&-&\frac{1}{2} \mathcal{T}_{\alpha\beta} \sqrt{-g} \delta
g^{\alpha\beta}d^ {4}x.
\end{eqnarray}
The explicit formulation of $\delta\mathcal{Q}$ is given in Appendix
\textbf{B}. Moreover, we define
\begin{eqnarray}\label{25}
\mathcal{T}_{\alpha\beta} \equiv \frac{-2}{\sqrt{-g}} \frac{\delta
(\sqrt{-g} \mathcal{L}_{M})}{\delta g^{\alpha\beta}}, \quad
\Theta_{\alpha\beta} \equiv g^{\gamma\eta} \frac{\delta
\mathcal{T}_{\gamma\eta}}{\delta g^{\alpha\beta}},
\end{eqnarray}
which implies that $ \delta \mathcal{T}= \delta
(\mathcal{T}_{\alpha\beta}g^{\alpha\beta}) =
(\mathcal{T}_{\alpha\beta}+ \Theta_{\alpha\beta})\delta
g^{\alpha\beta}$. Thus, Eq.(\ref{24}) turns out to be
\begin{eqnarray}\nonumber
\delta\mathcal{S} &=& \int \frac{1}{2}\bigg\{\frac{-1}{2}f
g_{\alpha\beta}\sqrt{-g} \delta g^{\alpha\beta} +
f_{\mathcal{T}}(\mathcal{T}_{\alpha\beta}+
\Theta_{\alpha\beta})\sqrt{-g} \delta g^{\alpha\beta}
\\\nonumber
&-&f_{\mathcal{Q}} \sqrt{-g} (\mathcal{P}_{\alpha\gamma\eta}
\mathcal{Q}_{\beta}^{\gamma\eta}- 2\mathcal{Q}^{\gamma\eta}
_{\alpha} \mathcal{P}_{\gamma\eta\beta}) \delta
g^{\alpha\beta}+2f_{\mathcal{Q}}\sqrt{-g}
\mathcal{P}_{\gamma\alpha\beta}\nabla^{\gamma} \delta
g^{\alpha\beta}
\\\label{26}
&-&\frac{1}{2} \mathcal{T}_{\alpha\beta}\sqrt{-g} \delta
g^{\alpha\beta}d^ {4}x.
\end{eqnarray}
Integrating and using the boundary conditions, the term $2
f_{\mathcal{Q}}\sqrt{-g}\mathcal{P}_{\gamma\alpha\beta}
\nabla^{\gamma}\delta g^{\alpha\beta}$ takes the form $-2
\nabla^{\gamma} (f_{\mathcal{Q}} \sqrt{-g}
\mathcal{P}_{\gamma\alpha\beta})\delta g^{\alpha\beta}$. Equating
the variation of Eq.(\ref{26}) to zero, we obtain the field
equations of $f(\mathcal{Q},\mathcal{T})$ theory as
\begin{eqnarray}\nonumber
\mathcal{T}_{\alpha\beta}&=& \frac{-2}{\sqrt{-g}} \nabla_{\gamma}
(f_{\mathcal{Q}}\sqrt{-g} \mathcal{P}^{\gamma}_{\alpha\beta})-
\frac{1}{2} f g_{\alpha\beta} + f_{\mathcal{T}}
(\mathcal{T}_{\alpha\beta} + \Theta_{\alpha\beta})
\\\label{27}
&-&f_{\mathcal{Q}} (\mathcal{P}_{\alpha\gamma\eta}
\mathcal{Q}_{\beta}^{\gamma\eta} -2\mathcal{Q}^{\gamma\eta}_{\alpha}
\mathcal{P}_{\gamma\eta\beta}),
\end{eqnarray}
where $f_{\mathcal{T}}$ represents the derivative corresponding to
the trace of energy-momentum tensor, whereas $f_{\mathcal{Q}}$
defines the derivative with respect to non-metricity. This
represents the modified field equations in the context of the
$f(\mathcal{Q},\mathcal{T})$ theory.

\section{Field Equations and Matching Conditions}

To explore the structure of CSs, we consider inner region as
\begin{equation}\label{28}
ds^{2}=dt^{2}e^{\mu(r)}- dr^{2}e^{\nu(r)}-d
\theta^{2}r^{2}-d\phi^{2}r^{2}\sin^{2}\theta.
\end{equation}
The stress-energy tensor manifests the configurations of matter and
energy in a system and its non-zero components yield physical
features. We consider anisotropic matter distribution as
\begin{equation}\label{29}
\mathcal{T}_{\alpha\beta}=\mathcal{U}_{\alpha}\mathcal{U}_{\beta}
\rho+P_{r}\mathcal{V}_{\alpha}\mathcal{V}_{\beta}-P_{t}
g_{\alpha\beta}+ P_{t}(\mathcal{U}_{\alpha}\mathcal{U}_{\beta}
-\mathcal{V}_{\alpha}\mathcal{V}_{\beta}),
\end{equation}
where four-vector and four-velocity of the fluid are denoted by
$\mathcal{V}_{\alpha}$ and $\mathcal{U}_{\alpha}$, respectively.

In the literature, one commonly considered matter Lagrangian density
for anisotropic matter is $\mathcal{L}_{m}=-\frac{P_{r}+2P_{t}}{3}$
\cite{41a}. The motivation for considering this matter-Lagrangian
lies in its ability to describe anisotropic matter configurations in
a simple and physically meaningful manner. For example, matter
distribution may exhibit anisotropic characteristics in various
astrophysical and cosmological scenarios. In the context of CSs, the
matter inside them might not be isotropic and different pressures
may act along different axes. The chosen form of the Lagrangian
density allows us to capture this anisotropy. The radial and
tangential pressures are physically meaningful quantities that are
often encountered in the study of anisotropic matter. This specific
form of the Lagrangian makes it easier to interpret the physical
significance of the pressure terms.

The components of $\Theta_{\alpha\beta}$ can be expressed as
\begin{eqnarray}\nonumber
\Theta_{11}&=& -\frac{1}{3}(6\rho+P_{r}+2P_{t})e^{\mu}, \quad
\Theta_{22}=-\frac{1}{3}(2P_{t}-5P_{r})e^{\nu } , \quad
\\\label{30}
\Theta_{33}&=& \frac{1}{3}(P_{r}-4P_{t})r^{2},\quad \Theta_{44}=
\frac{1}{3}(P_{r}-4P_{t})r^{2}\sin^{2} \theta.
\end{eqnarray}
By using the above constraints, we obtain the field equations of
$f(\mathcal{Q},\mathcal{T}$) gravity for static spherical spacetime
as
\begin{eqnarray}\nonumber
\rho&=&\frac{1}{2r^{2}e^{\nu}}\bigg[2r\mathcal{Q}'f_{\mathcal{Q}
\mathcal{Q}}(e^{\nu}-1)
+f_{\mathcal{Q}}\big((e^{\nu}-1)(2+r\mu')+(e^{\nu}+1)r\nu' \big)
\\\label{31}
&+&fr^{2}e^{\nu}\bigg]-\frac{1}{3}f_{\mathcal{T}}(3\rho+P_{r}+2P_{t}),
\\\nonumber
P_{r}&=&\frac{-1}{2r^{2}e^{\nu}}\bigg[2r\mathcal{Q}'f_{\mathcal{Q}
\mathcal{Q}}(e^{\nu}-1)
+f_{\mathcal{Q}}\big((e^{\nu}-1)(2+r\mu'+r\nu')-2r\mu'\big)
\\\label{32}
&+&fr^{2}e^{\nu}\bigg]+\frac{2}{3}f_{\mathcal{T}}(P_{t}-P_{r}),
\\\nonumber
P_{t}&=&\frac{-1}{4re^{\nu}}\bigg[-2r\mathcal{Q}'\mu'f_{\mathcal{Q}
\mathcal{Q}}
+f_{\mathcal{Q}}\big(2\mu'(e^{\nu}-2)-r\mu'^{2}+\nu'(2e^{\nu}+r\mu')
\\\label{33}
&-&2r\mu''\big)+2fre^{\nu}\bigg]+\frac{1}{3}f_{\mathcal{T}}
(P_{r}-P_{t}).
\end{eqnarray}
Now, we examine how $f(\mathcal{Q},\mathcal{T})$ affects the
geometry of CSs. We choose a specific model of
$f(\mathcal{Q},\mathcal{T})$ as \cite{42}
\begin{eqnarray}\label{34}
f(\mathcal{Q},\mathcal{T})=\xi\mathcal{Q}+\lambda\mathcal{T}.
\end{eqnarray}
This cosmological model has been widely used in the literature
\cite{42a}. The corresponding modified field equations lead to
\begin{eqnarray}\nonumber
\rho&=&\frac{\xi e^{-\nu}}{12r^2(2\lambda-1)(\lambda+1)}\bigg[
\lambda\big(2r(-\nu'(r\mu'+2)+2r\mu''+\mu'(r\mu'+4))-4e^{\nu}
\\\label{35}
&+&4\big)+3\lambda r(\mu'(-r\nu'+r\mu'+4)+2r\mu'')+12
(\lambda-1)(r\nu'+e^{\nu}-1)\bigg],
\\\nonumber
P_{r}&=&\frac{\xi e^{-\nu}}{12r^2(2\lambda-1)(\lambda+1)}\bigg[
2\lambda\big(r\nu'(r\mu'+2)+2(e^{\nu}-1)-r(2r\mu''+\mu'(r\mu'
\\\nonumber
&+&4))\big)+3\big(r\big(\lambda\nu'(r\mu'+4)-2\lambda
r\mu''-\mu'(-4\lambda+\lambda r\mu '+4)\big)-4(\lambda-1)
\\\label{36}
&\times&(e^{\nu}-1)\big)\bigg],
\\\nonumber
P_{t}&=&\frac{\xi e^{-\nu}}{12(2\lambda-1)r^2(\lambda+1)}\bigg[
2\lambda\big(r\nu'(r\mu'+2)+2(e^{\nu}-1)-r(2r\mu''+\mu'
\\\nonumber
&\times&(r\mu'+4))\big)+3\big(r
\big(2(\lambda-1)r\mu''-((\lambda-1)r\mu'-2)(\nu'-\mu')\big)
\\\label{37}
&+&4\lambda(e^{\nu}-1)\big)\bigg].
\end{eqnarray}

Recently, Krori-Barua solutions have gained much attention because
of their non-singular behavior, defined as \cite{43}
\begin{eqnarray}\label{38}
e^{\mu(r)}=e^{br^{2}+c}, \quad e^{\nu(r)}=e^{ar^{2}},
\end{eqnarray}
where arbitrary constants are denoted by $a$, $b$ and $c$. The
observed values of mass and radius of the considered stars are given
in Table \textbf{1}, while the constants corresponding to mass and
radius are shown in Table \textbf{2}. The compatibility of the
solution is ensured by the non-singular and positively increasing
behavior of metric elements throughout the domain. The behavior of
these metric potentials is demonstrated in Figure \textbf{1} which
manifests that both metric elements are regular and show positively
increasing behavior as required. In all graphs, we use CS1, CS2,
CS3, CS4, CS5, CS6, CS7, CS8 for 4U 1538-52, LMC X-4, Cen X-3, 4U
1608-52, PSR J1903+327, PSR J1614-2230, Vela X-1 and SMC X-4 CSs,
respectively.
\begin{table}\caption{Approximate values of input parameters.}
\begin{center}
\begin{tabular}{|c|c|c|c|c|c|c|c|}
\hline Compact stars & $M_{\odot}$ & $h(km)$
\\
\hline  4U 1538-52 \cite{43a} & 0.87 $\pm$ 0.07 & 7.866 $\pm$ 0.21
\\
\hline  LMC X-4 \cite{43a} & 1.04 $\pm$ 0.09 & 8.301 $\pm$ 0.2
\\
\hline  Cen X-3 \cite{43a} & 1.49 $\pm$ 0.08 & 9.178 $\pm$ 0.13
\\
\hline  4U 1608-52 \cite{43b} & 1.74 $\pm$ 0.01 & 9.3 $\pm$ 0.10
\\
\hline  PSR J1903+327 \cite{43c} & 1.667 $\pm$ 0.021 & 9.48 $\pm$
0.03
\\
\hline  PSR J1614-2230 \cite{43d} & 1.97 $\pm$ 0.04 & 9.69 $\pm$ 0.2
\\
\hline  Vela X-1 \cite{43a} & 1.77 $\pm$ 0.08 & 9.56 $\pm$ 0.08
\\
\hline  SMC X-4  \cite{43a} & 1.29 $\pm$ 0.05 & 8.831 $\pm$ 0.09
\\
\hline
\end{tabular}
\end{center}
\end{table}
\begin{table}\caption{Approximate values of output parameters.}
\begin{center}
\begin{tabular}{|c|c|c|c|c|c|c|c|}
\hline Compact stars & $\mathrm{a}$ & $\mathrm{b}$  & $\mathrm{c}$
\\
\hline  4U 1538-52 & 0.00637763 & 0.00390959 & -0.636511
\\
\hline  LMC X-4 & 0.00669013 & 0.00424959 & -0.753818
\\
\hline  Cen X-3 & 0.00773096 & 0.00544831 & -1.11017
\\
\hline  4U 1608-52 & 0.00927255 & 0.00711041 & -1.41696
\\
\hline  PSR J1903+327 & 0.00812966 & 0.0059884 & -1.2688
\\
\hline  PSR J1614-2230 & 0.00974088 & 0.00796547 & -1.66256
\\
\hline  Vela X-1 & 0.00863565 & 0.00657448 & -1.39011
\\
\hline  SMC X-4 & 0.00722214 & 0.00484915 & -0.941398
\\
\hline
\end{tabular}
\end{center}
\end{table}
\begin{figure}
\epsfig{file=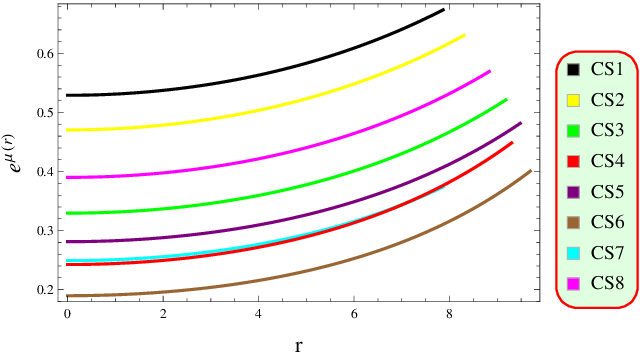,width=.5\linewidth}
\epsfig{file=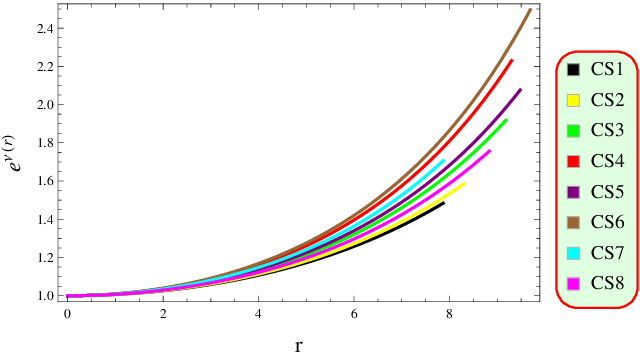,width=.5\linewidth}\caption{Graphs of metric
potentials versus radial coordinate.}
\end{figure}

The unknown constants $(a,b,c)$ can be manipulated by using the
first Darmois junction condition. This condition is used to describe
how different regions of spacetime can be smoothly connected at a
boundary. This constraint provides a way to match two different
solutions of the field equations across a hypersurface, which is
often used to model situations where one region represents an
interior solution and the other region represents an exterior
solution. The first fundamental form of Darmois junction conditions
(continuity of metric potentials) states that the metric potentials
should be continuous across the boundary that separates the inner
and outer regions. This is important to ensure a smooth transition
between the interior and exterior solutions, maintaining the
integrity of the spacetime geometry.

We consider the outer geometry of CSs as
\begin{eqnarray}\label{39}
ds^{2}_{+}=-\frac{1}{r}(r-2m)dt^{2}+r(r-2m)^{-1}d
r^{2}+r^{2}(d\theta^{2} +\sin^{2}\theta d\phi^{2}),
\end{eqnarray}
where $m$ represents the total mass of the outer geometry. The
continuity of metric coefficients of the metrics (\ref{28}) and
(\ref{39}) at the surface boundary $(r=h)$ gives
\begin{eqnarray}\nonumber
g_{tt}=e^{bh^{2}+c}=1-\frac{2m}{h}, \quad
g_{rr}=e^{-ch^{2}}=1-\frac{2m}{h}, \quad
g_{tt}=bhe^{bh^{2}+c}=\frac{m}{h^{2}}.
\end{eqnarray}
By solving the above equations simultaneously, we obtain
\begin{eqnarray}\label{40}
a=-\frac{1}{h^2}\ln(1-\frac{2m}{h}), \quad b=\frac{m}{h^2(h-2m)},
\quad c=\frac{m}{2m-h}\ln(1-\frac{2m}{h}).
\end{eqnarray}
These constraints are important to comprehended hidden aspects of
the CSs. The corresponding field equations are
\begin{eqnarray}\nonumber
\rho&=&\frac{\xi e^{-ar^2}}{3(2\lambda^2+\lambda-1)r^2}\bigg[r^2
(a(-5b\lambda r^2+4\lambda-6)+5b\lambda(br^2+3))-2\lambda+3
\\\label{41}
&+&(2\lambda-3)e^{ar^2}\bigg],
\\\nonumber
P_{r}&=&\frac{\xi e^{-a r^2}}{3(2\lambda^2+\lambda-1)r^2}\bigg[r^2(8
a\lambda-b(\lambda(5r^2(b-a)+3)+6))+(3-2\lambda)e^{ar^2}
\\\label{42}
&+&2\lambda-3\bigg],
\\\nonumber
P_{t}&=&\frac{\xi e^{-ar^2}}{3(2\lambda^2+\lambda-1)r^2}\bigg[r^2
(a(-b(\lambda-3)r^2+2\lambda+3)+b(b(\lambda-3)r^2
\\\label{43}
&-&3(\lambda +2)))+4 \lambda(e^{ar^2}-1)\bigg].
\end{eqnarray}

\section{Physical Characteristics of Compact Stars}

We analyze viable characteristics of CSs and examine their behavior
graphically in this section. The following regularity constraints
inside the stellar objects must be satisfied for viable and stable
CSs with a certain radius.
\begin{itemize}
\item
The metric functions should be finite and non-singular, ensuring
that the spacetime is smooth and free from singularities.
\item
The positive and maximum behavior of matter contents at the center
of the CSs ensures that it has a stable core and decreases towards
the boundary, making the CSs physically viable. Moreover, the radial
pressure should vanish at the surface boundary, i.e.,
$P_{r}(r=h)=0$.
\item
The gradient of matter contents must vanish at the center and then
show negative behavior towards the boundary.
\item
The pressure components must be equal at $r=0$, which demonstrates
the anisotropy vanishing at the center of CSs. The positive behavior
of anisotropy indicates that pressure is directed outward and the
negative behavior implies that pressure is in the inward direction.
\item
The energy constraints must be positive to ensure the presence of
ordinary matter, which is necessary to obtain viable CSs.
\item
The EoS parameters should satisfy the range
$0\leq\omega_{r},\omega_{t}\leq1$.
\item
The compactness is a dimensionless quantity which is used to examine
the viability of compact stars. The compactness factor must be less
than 4/9 for viable stellar structures.
\item
The redshift function measures the force exerted on light by strong
gravity which validates the physical existence of stellar objects.
The redshift should be less than or equal to 5.2 for viable
anisotropic compact objects.
\item
All forces (gravitational, hydrostatic and anisotropic) must satisfy
the equilibrium condition.
\item
The sound speed components must lie in [0,1], i.e., $0<u_{sr},
u_{st}<1$, which is important to maintain a model stable.
\item
An anisotropic fluid sphere must have an adiabatic index greater
than 4/3.
\end{itemize}
We examine the impact of different physical parameters such as
matter variables, anisotropy, energy bounds, EoS parameters, mass,
compactness, redshift, equilibrium state (TOV equation), and
stability analysis (sound speed and adiabatic index) through graphs.

\subsection{Evolution of Matter Contents}

The graphical behavior of fluid parameters and their derivatives for
each star candidate is shown in Figures \textbf{2}-\textbf{3}. It is
found that these physical characteristics are maximum at the center
and positively decreasing, revealing a highly compact profile of the
proposed CSs. Moreover, the radial pressure inside each candidate
shows monotonically decreasing behavior with the rise in $r$ and
vanishes at the boundary. Figure \textbf{3} shows that the
derivative of fluid parameters is zero at the center and negative,
which confirms the existence of highly compact configuration in
$f(\mathcal{Q},\mathcal{T})$ theory. The graphical behavior shows
that fluid parameters have greater values than GR \cite{44}.
\begin{figure}
\epsfig{file=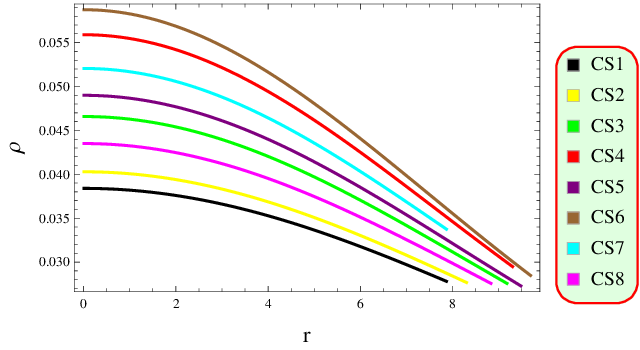,width=.5\linewidth}
\epsfig{file=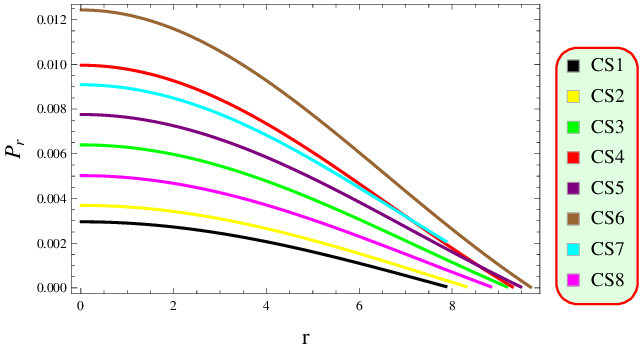,width=.5\linewidth}\center
\epsfig{file=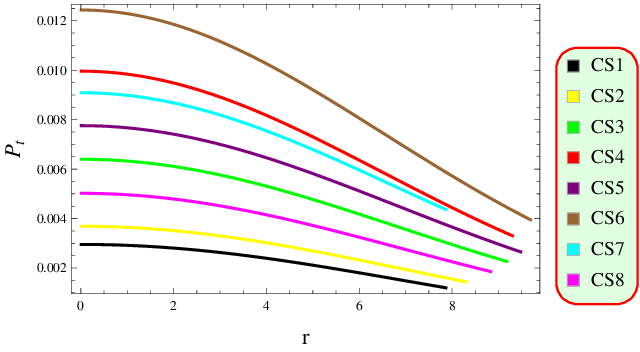,width=.5\linewidth}\caption{Evolution of matter
contents versus radial coordinate for $\xi=2$ and $\lambda=-0.005$.}
\end{figure}
\begin{figure}
\epsfig{file=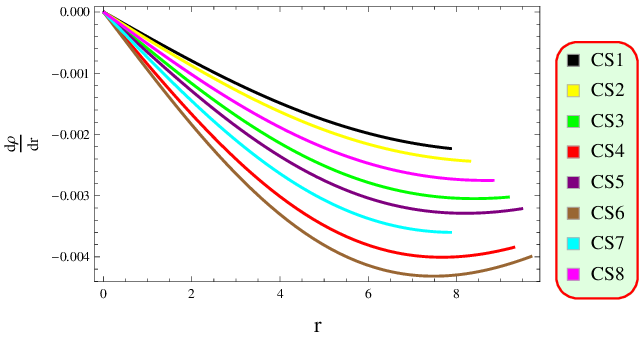,width=.5\linewidth}
\epsfig{file=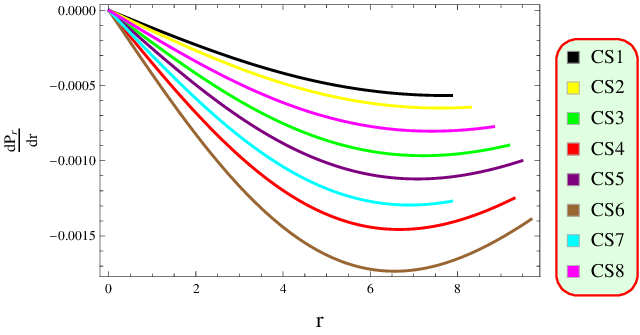,width=.5\linewidth}\center
\epsfig{file=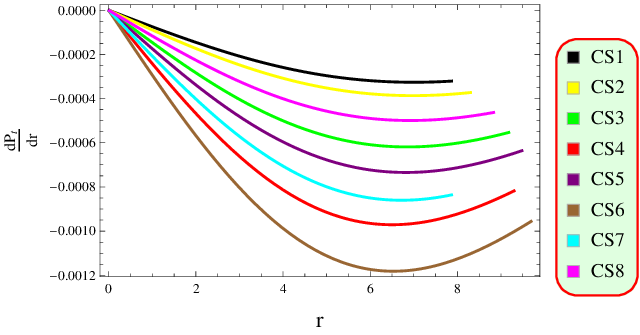,width=.5\linewidth}\caption{Evolution of
gradient of matter contents versus radial coordinate for $\xi=2$ and
$\lambda=-0.005$.}
\end{figure}
\begin{figure}\center
\epsfig{file=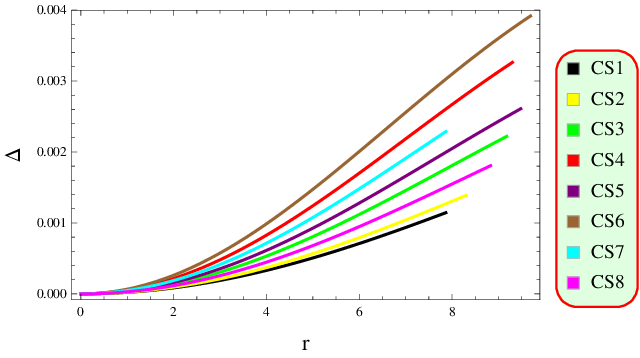,width=.5\linewidth}\caption{Behavior of
$P_t-P_r$ versus radial coordinate for $\xi=2$ and
$\lambda=-0.005$.}
\end{figure}

\subsection{Anisotropic Pressure}

The pressure anisotropy $(\Delta=P_t-P_r)$ refers to the phenomenon
where the pressure in a system is not equal in all directions. The
behavior of anisotropy for the considered CSs is given in Figure
\textbf{4}. It is found that anisotropy shows positively increasing
behavior for all CSs, which ensures the existence of repulsive force
that is necessary for massive geometries \cite{45}. Moreover, the
anisotropy in this theory increases in contrast to GR \cite{46}.

\subsection{Energy Conditions}

In order to investigate the existence of viable cosmic structures,
it is necessary to apply some specific constraints on matter named
as energy conditions. These conditions consist of a set of
inequalities that impose limitations on the stress-energy tensor
which governs the behavior of matter and energy in the presence of
gravity.
\begin{itemize}
\item \textbf{Null energy constraint}\\
According to this condition, the energy density observed by any
observer moving at the speed of light cannot be negative. This is
defined as
\begin{eqnarray}\nonumber
0\leq P_{r}+\rho, \quad 0\leq P_{t}+\rho.
\end{eqnarray}
\item \textbf{Dominant energy constraint}\\
This determines that the energy density must be greater than or
equal to the energy flux as measured by any observer.
Mathematically, it is expressed as
\begin{eqnarray}\nonumber
0\leq \rho\pm P_{r}, \quad 0\leq \rho\pm P_{t}.
\end{eqnarray}
\item \textbf{Weak energy constraint}\\
This constraint implies that the energy density measured by an
observer is non-negative. Also, the sum of energy density and
pressure components must be non-negative, expressed as
\begin{eqnarray}\nonumber
0\leq P_{r}+\rho,\quad 0\leq P_{t}+\rho, \quad 0\leq \rho.
\end{eqnarray}
\item \textbf{Strong energy constraint}\\
This condition is a stronger version of the weak energy constraint
and states that not only the energy density is non-negative but the
addition of $\rho, P_r, P_{t}$ is also non-negative. This can be
represented as
\begin{eqnarray}\nonumber
0\leq P_{r}+\rho, \quad 0\leq P_{t}+\rho, \quad 0\leq
P_{r}+2P_{t}+\rho.
\end{eqnarray}
\end{itemize}
These energy bounds have a significant impact on the existence of
viable cosmic objects in spacetime. The viable cosmic structure must
satisfy these conditions. Figure \textbf{5} demonstrates that matter
inside the CSs is ordinary as all the energy constraints are
satisfied in the presence of $f(\mathcal{Q},\mathcal{T})$ terms.
\begin{figure}
\epsfig{file=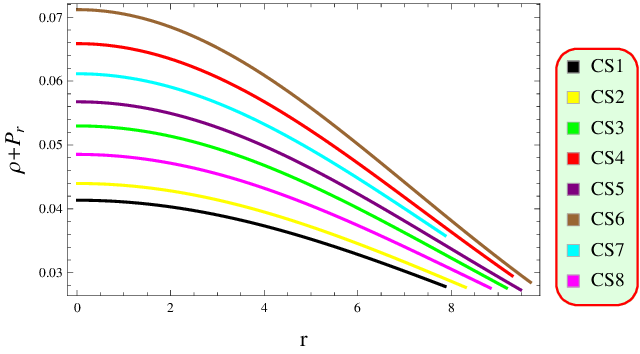,width=.5\linewidth}
\epsfig{file=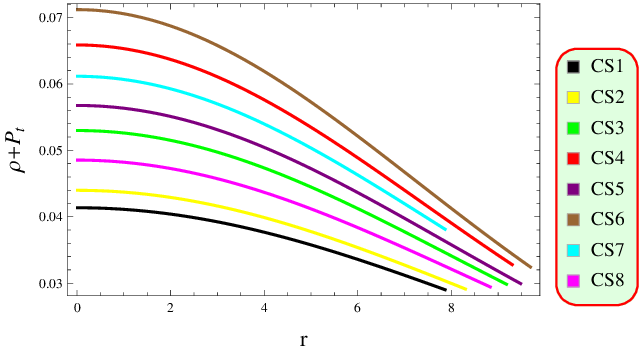,width=.5\linewidth}
\epsfig{file=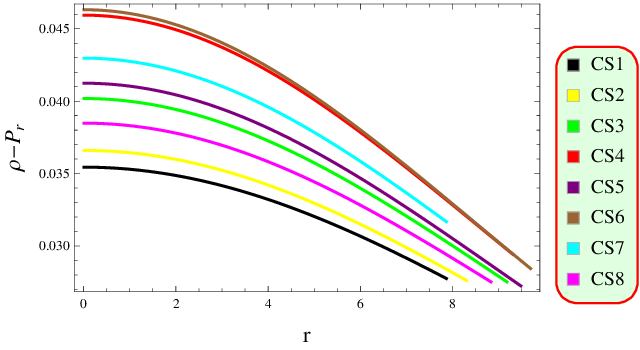,width=.5\linewidth}
\epsfig{file=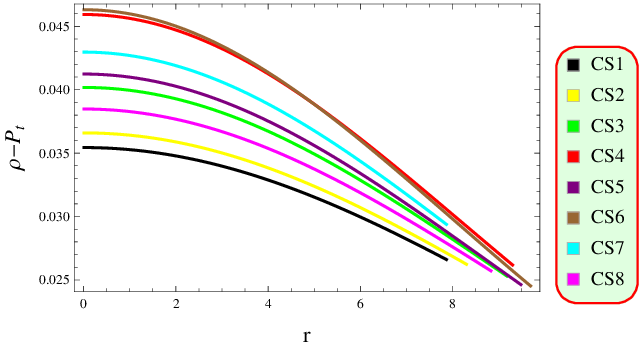,width=.5\linewidth}\caption{Graphs of energy
conditions versus radial coordinate for $\xi=2$ and
$\lambda=-0.005$.}
\end{figure}

\subsection{Equation of State Parameters}

Here, we investigate the EoS parameters that are crucial in
describing the relation between pressure and energy density in
various physical systems. For a physically viable model, the radial
$(\omega_{r}=\frac{P_{r}}{\rho})$ and transverse
$(\omega_{t}=\frac{P_{t}}{\rho})$ EoS parameters must lie in [0,1]
\cite{47}. Using Eqs.(\ref{41})-(\ref{43}), we have
\begin{eqnarray}\nonumber
\omega_{r}&=&-1-\big(6(2\lambda-1)r^2(a+b)\big)\bigg[r^2\big(a(5b\lambda
r^2-4\lambda+6)-5b\lambda (br^2+3)\big)
\\\nonumber
&+&(3-2\lambda)e^{ar^2}+2\lambda-3\bigg]^{-1},
\\\nonumber
\omega_{t}&=&\bigg[r^2\big(a(-b(\lambda-3)r^2+2\lambda+3)+b(b(\lambda-3)
r^2-3(\lambda+2))\big)+4\lambda
\\\nonumber
&\times&(e^{ar^2}-1)\bigg]\bigg[r^2\big(a(-5b\lambda r^2+4
\lambda-6)+5b\lambda(br^2+3)\big)+(2\lambda-3)e^{ar^2}
\\\nonumber
&-&2\lambda+3\bigg]^{-1}.
\end{eqnarray}
The graphical analysis of EoS parameters is given in Figure
\textbf{6}, which shows that $\omega_{r}$ and $\omega_{t}$ satisfy
the required viability condition of the considered CSs. Moreover,
the range of EoS parameters are maximum than GR \cite{46}.
\begin{figure}
\epsfig{file=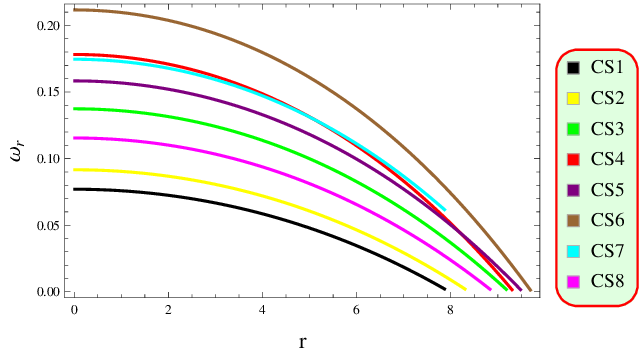,width=.5\linewidth}
\epsfig{file=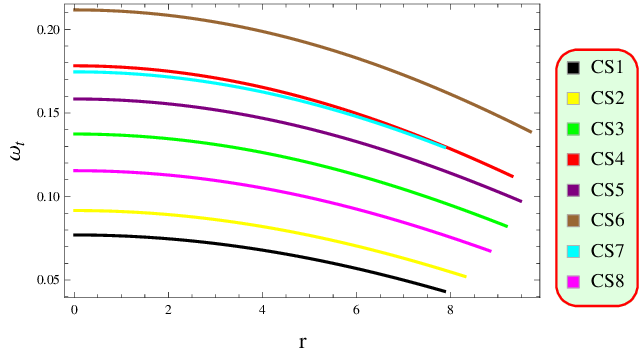,width=.5\linewidth}\caption{Graphs of radial
and tangential components of EoS parameter versus radial coordinate
for $\xi=2$ and $\lambda=-0.005$.}
\end{figure}

\subsection{Mass, Compactness and Redshift}

The mass of CS is defined as
\begin{equation}\label{44}
M=4\pi\int^{h}_{0} r^{2}\rho dr.
\end{equation}
The numerical solution of this equation for our considered model is
obtained using an initial condition $M(0)=0$. We examine the
graphical behavior of mass inside each CS resulting from this
numerical solution in Figure \textbf{7}, which manifests that the
mass increases positively and monotonically as the radius increases.
Also, $M\rightarrow 0$ as $r\rightarrow 0$ which shows that the mass
function is regular at the center of CSs. The compactness function
$(u=\frac{M}{r})$ is one of them which plays a crucial role in
examining the viability of the CSs. Buchdahl \cite{48} proposed a
specified limit of mass-radius ratio as $u<4/9$ for viable CSs.

The surface redshift measures the change in the wavelength of light
emitted from the surface of a CS due to the strong gravitational
influence of the object. This can be expressed in terms of
compactness as
\begin{equation}\label{45}
Z_s =- 1+\frac{1}{\sqrt{1-2u}}.
\end{equation}
Buchdahl \cite{48} established that the value of surface redshift
must be less than 2 for viable CSs with perfect matter distribution,
but Ivanov \cite{49} detected a value of 5.211 for anisotropic
configurations when the dominant energy condition holds. The
behavior of both compactness and redshift functions is monotonically
increasing and vanishing at the center of the star, as shown in
Figure \textbf{8}. Further, both functions lie in the specified
limits ($u<4/9$ and $Z_{s}<5.211$).
\begin{figure}\center
\epsfig{file=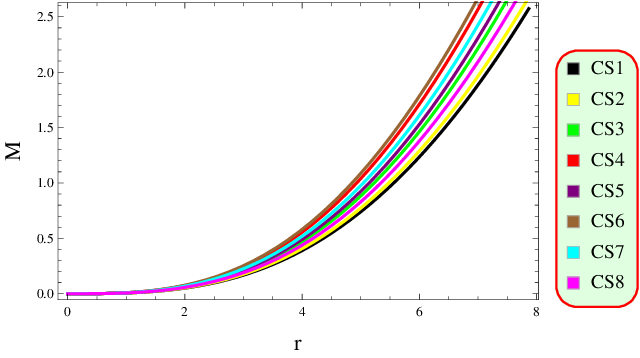,width=.5\linewidth} \caption{Plots of mass
versus radial coordinate for $\xi=2$ and $\lambda=-0.005$.}
\end{figure}
\begin{figure}
\epsfig{file=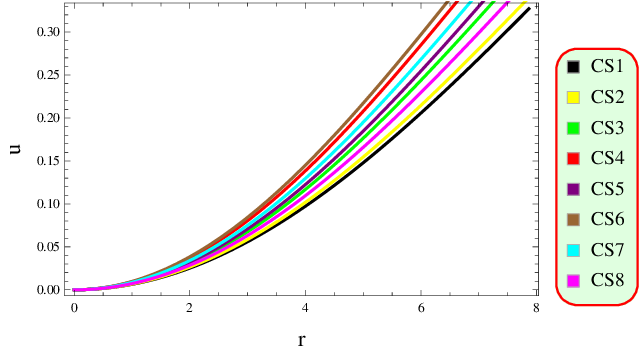,width=.5\linewidth}
\epsfig{file=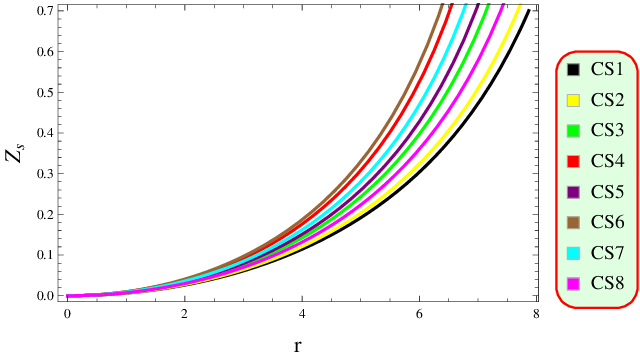,width=.5\linewidth}\caption{Graph of
compactness factor and surface redshift versus radial coordinate for
$\xi=2$ and $\lambda=-0.005$.}
\end{figure}

\section{Equilibrium and Stability Analysis}

Equilibrium state and stability analysis are essential concepts in
understanding the structure and behavior of cosmic objects. An
equilibrium state is a state of balance in which the internal and
external forces acting on the CSs are in equilibrium. Stability is
used to investigate the conditions under which cosmic structures
remain stable against various modes of oscillations. Here, we use
the \emph{sound speed} and \emph{adiabatic index} methods to analyze
the stability of CSs. Sound speed defines the rate at which pressure
waves propagate through a medium, while the adiabatic index
characterizes the relationship between pressure and density changes
in the CSs.

\subsection{Tolman-Oppenheimer-Volkoff Equation}

This fundamental equation in astrophysics describes the equilibrium
structure of the static spherically symmetric spacetime. It gives
information how the star's pressure and gravitational forces are
balanced to maintain its equilibrium. The TOV equation for
anisotropic matter configuration is \cite{50}
\begin{equation}\label{46}
\frac{M_{G}(r)e^\frac{\mu-\nu}{2}}{r^{2}}(\rho+P_{r})+
\frac{dP_{r}}{dr}-\frac{2}{r}(P_{t}-P_{r})=0,
\end{equation}
where the gravitational mass is determined as
\begin{equation}\nonumber
M_{G}(r)=4\pi \int (\mathcal{T}^{t}_{t} - \mathcal{T}^{r}_{r} -
\mathcal{T}^{\theta}_{\theta} - \mathcal{T}^{\phi}_{\phi}) r^{2}
e^{\frac{\mu+\nu}{2}}dr.
\end{equation}
Solving this equation, we have
\begin{equation}\nonumber
M_{G}(r) = \frac{1}{2}r^{2} e^{\frac{\nu-\mu}{2}}\mu'.
\end{equation}
Substituting this value in Equation (\ref{46}), we obaian
\begin{equation}\nonumber
\frac{1}{2}\mu'(\rho+P_{r})+
\frac{dP_{r}}{dr}-\frac{2}{r}(P_{t}-P_{r})=0.
\end{equation}
This describes how the pressure gradient changes with radial
distance inside the star. The solution of the TOV equation provides
information about the internal structure of the stars such as its
density profile and pressure distribution. This demonstrates the
influence of gravitational
$\big(F_{g}=\frac{\mu'(\rho+P_{r})}{2}\big)$, hydrostatic
$\big(F_{h}=\frac{dP_{r}}{dr}\big)$ and anisotropic
$\big(F_{a}=\frac{2(P_{r}-P_{t})}{r}\big)$ forces on the system.
Using Eqs.(\ref{41})-(\ref{43}), we obtain
\begin{eqnarray}\nonumber
F_{g}&=&\frac{2\xi  br(a+b)e^{-ar^2}}{\lambda+1}, \quad F_{a}=
-\frac{2\xi e^{-ar^2}(br^4(b-a)-ar^2+e^{a r^2}-1)}{(\lambda+1)r^3},
\\\nonumber
F_{h}&=&\frac{2\xi e^{-ar^2}}{3(2\lambda^2+\lambda-1)r^3}\bigg[
-\lambda\big(r^4(8a^2-8ab+5b^2)+5abr^6(a-b)+2ar^2+2\big)
\\\nonumber
&+&3a(2b r^4+r^2)+(2\lambda-3)e^{ar^2}+3\bigg].
\end{eqnarray}
Figures \textbf{9} shows that the our considered CSs are in
equilibrium state as the total effect of $F_{g}$, $F_{h}$ and
$F_{a}$ is zero.
\begin{figure}\center
\epsfig{file=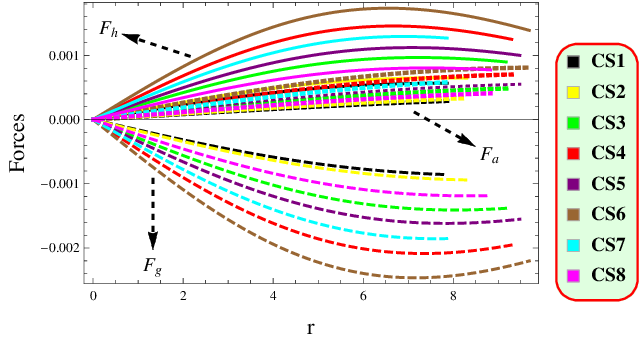,width=.5\linewidth}\caption{Plot of TOV
equation for $\xi=2$ and $\lambda=-0.005$.}
\end{figure}

\subsection{Casuality Condition}

The causality condition states that the time-like interval between
any two events in spacetime must always be greater than or equal to
zero, i.e., no signal can travel faster than the speed of light.
According to this condition, the radial and tangential components of
sound speed $(u_{sr}=\frac{dP_{r}}{d\rho}, u_{st}
=\frac{dP_{t}}{d\rho})$ must lie in [0,1] interval for stable
structures \cite{51}. The sound speed's components in the framework
of $f(\mathcal{Q},\mathcal{T})$ are
\begin{eqnarray}\nonumber
u_{sr}&=&-1-(6a(2\lambda-1)r^4(a+b))\bigg[a^2r^4(5b\lambda
r^2-4\lambda +6)-ar^2(\lambda(5br^2(b r^2
\\\nonumber
&+&4)-2)+3)+(3-2\lambda)e^{ar^2}+5b^2\lambda
r^4+2\lambda-3\bigg]^{-1},
\\\nonumber
u_{st}&=&\bigg[a^2r^4(b(\lambda-3)r^2-2\lambda-3)+a
\big(-b^2(\lambda-3)r^6+b(2\lambda+9)r^4+4\lambda r^2\big)
\\\nonumber
&-&-4\lambda(e^{ar^2}-1)+b^2(\lambda-3)r^4\bigg]
\bigg[a^2r^4(5b\lambda r^2-4\lambda+6)-ar^2 \big(\lambda(5br^2
\\\nonumber
&\times&(br^2+4)-2)+3\big)+(3-2 \lambda)e^{ar^2}+5b^2\lambda
r^4+2\lambda-3\bigg]^{-1}.
\end{eqnarray}
Figure \textbf{10} shows that static spherically symmetric solutions
are in the stable state as they fulfill the necessary constraints.
Thus, physically viable and stable CSs exist in this modified
theory.
\begin{figure}
\epsfig{file=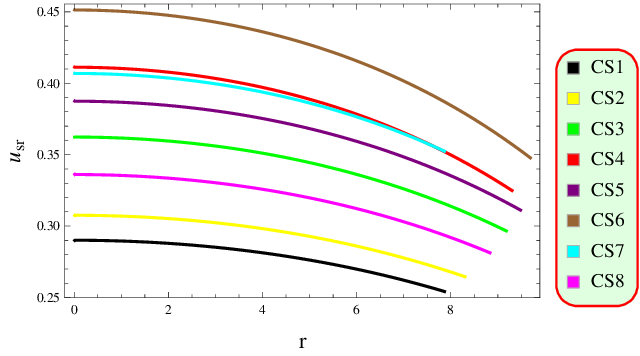,width=.5\linewidth}
\epsfig{file=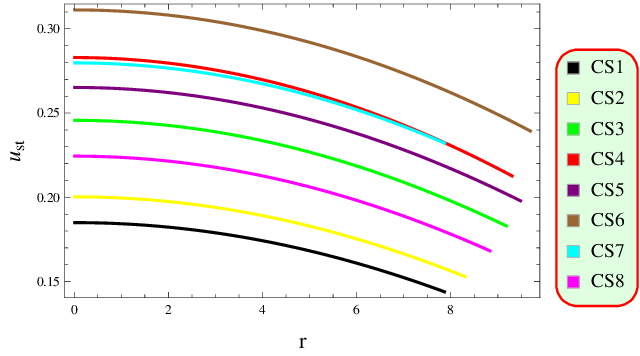,width=.5\linewidth}\caption{Plots of sound
speed for $\xi=2$ and $\lambda=-0.005$.}
\end{figure}

\subsection{Herrera Cracking Approach}

The stability of a CSs can be determined by analyzing the behavior
of the cracking condition $(0\leq\mid u_{st}-u_{sr}\mid\leq1)$
\cite{52}. If the cracking condition is violated, then the CSs are
unstable and will collapse while if the cracking condition is
satisfied, then the CSs are stable and can exist for a long time.
Figure \textbf{11} determines that considered CSs are stable as they
lie in the specified limit.
\begin{figure}\center
\epsfig{file=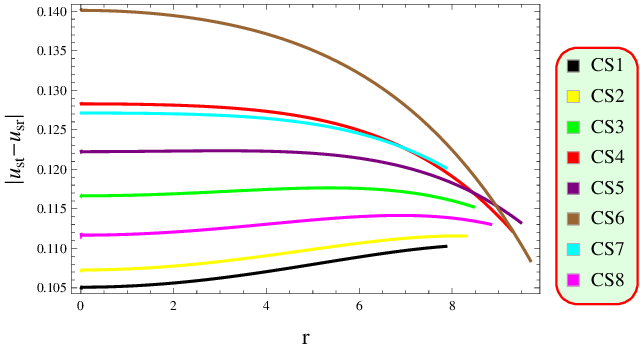,width=.5\linewidth}\caption{Behavior of Herrera
cracking approach for $\xi=2$ and $\lambda=-0.005$.}
\end{figure}

\subsection{Adiabatic Index}

Another technique for determining the stability of CSs is the
adiabatic index. The adiabatic index is defined as
\begin{eqnarray}\nonumber
\Gamma_{r}=\frac{\rho+P_{r}}{P_{r}} v_{sr},\quad
\Gamma_{t}=\frac{\rho+P_{t}}{P_{t}} v_{st},
\end{eqnarray}
where $\Gamma_{r}$ and $\Gamma_{t}$ are the radial and tangential
components of adiabatic index. Using Eqs.(\ref{41})-(\ref{43}), the
above equations become
\begin{eqnarray}\nonumber
\Gamma_{r}&=&-\bigg[6(2\lambda-1)r^2(a+b)\big(a^2\lambda r^4(5b
r^2+8)-ar^2(br^2(5b\lambda r^2+8\lambda+6)
\\\nonumber
&-&2\lambda +3)+(3-2\lambda)e^{ar^2}+5b^2\lambda
r^4+2\lambda-3\big)\bigg] \bigg[\big(a^2r^4(-5b\lambda r^2+4\lambda
\\\nonumber
&-&6)+ar^2\big(\lambda(5br^2(br^2+4)-2)+3\big)+(2\lambda-3)e^{ar^2}-5
b^2\lambda r^4-2\lambda
\\\nonumber
&+&3\big)\big(r^2\big(b(\lambda(5r^2(b-a)+3)+6)-8a\lambda\big)+(2\lambda
-3)e^{ar^2}-2\lambda+3\big)\bigg]^{-1},
\\\nonumber
\Gamma_{t}&=&\bigg[3(2\lambda-1)(br^4(b-a)+r^2(a+2b)+e^{a
r^2}-1)\big(-\lambda r^2\big(r^2(-2a^2+b^2
\\\nonumber
&+&2ab)+abr^4(a-b)+4a\big)+3r^4(a^2+abr^2(a-b)-3ab+b^2)+4\lambda
\\\nonumber
&\times&(e^{ar^2}-1)\big)\bigg]\bigg[\big(a^2r^4(5b\lambda
r^2-4\lambda +6)-ar^2\big(\lambda(5br^2(br^2+4)-2)+3\big)
\\\nonumber
&+&(3-2\lambda)e^{ar^2}+5b^2\lambda
r^4+2\lambda-3\big)\big(r^2\big(a
(b(\lambda-3)r^2-2\lambda-3)+b(3(\lambda
\\\nonumber
&+&2)-b(\lambda-3)r^2)\big)-4\lambda(e^{ar^2}-1)\big)\bigg]^{-1}.
\end{eqnarray}
If the value of $\Gamma$ is greater than 4/3 then CSs are stable,
otherwise CSs are unstable and will collapse \cite{53}. Figure
\textbf{12} shows that our system is stable in the presence of
correction terms as it satisfied the required limit. Hence, we
obtain viable and stable CSs in $f(\mathcal{Q},\mathcal{T})$ theory.
\begin{figure}
\epsfig{file=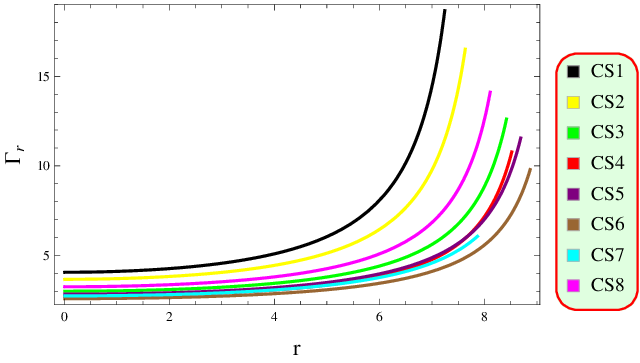,width=.5\linewidth}
\epsfig{file=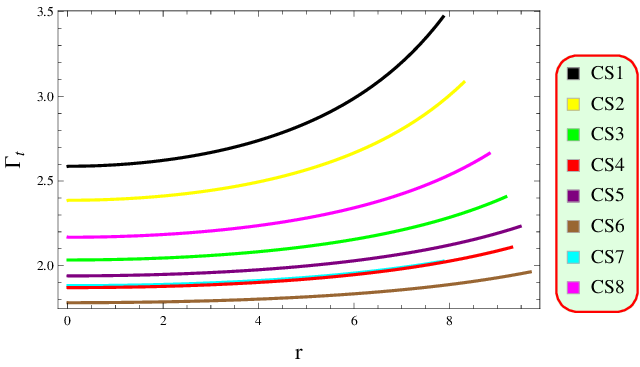,width=.5\linewidth}\caption{Behavior of
adiabatic index for $\xi=2$ and $\lambda=-0.005$.}
\end{figure}

\section{Final Remarks}

In this paper, we have examined the viability and stability of CSs
in $f(\mathcal{Q},\mathcal{T})$ theory. The main results are given
as follows.
\begin{itemize}
\item
We have found that both metric elements (Figure \textbf{1}) are
consistent and fulfill the necessary conditions, i.e., they exhibit
minimum value at the center of stars and then show monotonically
increasing behavior.
\item
The behavior of fluid parameters (Figure \textbf{2}) is positive and
regular in the interior of CSs and diminish at the boundary. Also,
the derivative of fluid parameters (Figure \textbf{3}) is negative
which presents a dense picture of the CSs.
\item
We have found that the anisotropic pressure (Figure \textbf{4}) is
directed outward which is necessary for compact stellar
configuration.
\item
All energy bounds are satisfied to confirm the presence of normal
matter in the interior of CSs (Figure \textbf{5}).
\item
The range of EoS parameters (Figure \textbf{6}) lie between 0 and 1,
which shows the viability of the considered model.
\item
We have found that the mass function is regular at the center of the
star $(\lim_{r\to 0}M=0)$ and show monotonically increasing behavior
as the radial coordinate increases (Figure \textbf{7}). The
compactness and redshift functions satisfy the required conditions
(Figure \textbf{8}).
\item
The TOV equation shows that gravitational, hydrostatic, and
anisotropic forces have a null impact for all proposed CSs (Figure
\textbf{9}). This suggests that the compact stellar models are in an
equilibrium state.
\item
The stability limits, i.e., $u_{sr}$ and $u_{st}\in[0,1]$ (causality
condition), $0<|u_{st}- u_{sr}|<1$ (Herrera cracking) approach and
$\Gamma>4/3$ (adiabatic index) are satisfied, which ensures the
existence of physically stable CSs (Figures
\textbf{10}-\textbf{12}).
\end{itemize}

We have obtained a more dense profile of the CSs through a
comprehensive analysis of the resulting solutions. It is interesting
to note that the range of physical quantities in this modified
gravity increases and provides more viable and stable CSs than GR
\cite{44}-\cite{46} and other modified theories \cite{54}-\cite{56}.
In $f(\mathcal{R})$ theory, it is found that the Her X-1 CS
corresponding to the second gravity model is not stable only
satisfying a very small range \cite{57}. It has been observed that
CSs are not physically viable and stable at the center in
$f(\mathcal{R},\mathcal{T}^{2})$ theory \cite{58}. Here, we have
found that all the considered CSs are physically viable and stable
in this modified theory.

\vspace{0.25cm}

\section*{Appendix A: Non-Metricity Scalar}
\renewcommand{\theequation}{A\arabic{equation}}
\setcounter{equation}{0}

According to Eqs.(\ref{19}) and (\ref{20}), we have
\begin{eqnarray}\nonumber
\mathcal{Q} &\equiv& -g^{\alpha\beta}
(\mathcal{L}^{\gamma}_{\eta\alpha}\mathcal{L}^{\eta}_{\beta\gamma} -
\mathcal{L}^{\gamma}_{\eta\gamma}\mathcal{L}^{\eta}_{\alpha\beta}),
\\\nonumber
\mathcal{L}^{\gamma}_{\eta\alpha}&=&-\frac{1}{2}
g^{\gamma\varsigma}(\mathcal{Q}_{\alpha\eta\varsigma}
+\mathcal{Q}_{\eta\varsigma\alpha}-\mathcal{Q}_{\varsigma\alpha\eta}),
\\\nonumber
\mathcal{L}^{\eta}_{\beta\gamma}&=&-\frac{1}{2}
g^{\eta\varsigma}(\mathcal{Q}_{\gamma\beta\varsigma}
+\mathcal{Q}_{\beta\varsigma\gamma}-\mathcal{Q}_{\varsigma\gamma\beta}),
\\\nonumber
\mathcal{L}^{\gamma}_{\eta\alpha} &=& -\frac{1}{2}
g^{\gamma\varsigma}(\mathcal{Q}_{\gamma\eta\varsigma}
+\mathcal{Q}_{\eta\varsigma\gamma}-\mathcal{Q}_{\varsigma\gamma\eta}),
\\\nonumber
&=&-\frac{1}{2}(\tilde{\mathcal{Q}_{\eta}}
+\mathcal{Q}_{\eta}-\tilde{\mathcal{Q}_{\eta}})= \quad -\frac{1}{2}
\mathcal{Q}_{\eta},
\\\nonumber
\mathcal{L}^{\eta}_{\alpha\beta}&=&
-\frac{1}{2}g^{\eta\varsigma}(\mathcal{Q}_{\beta\alpha\varsigma}
+\mathcal{Q}_{\alpha\varsigma\beta}-\mathcal{Q}_{\varsigma\beta\alpha}).
\end{eqnarray}
Thus, we have
\begin{eqnarray}\nonumber
-g^{\alpha\beta}\mathcal{L}^{\gamma}_{\eta\alpha}\mathcal{L}^{\eta}_{\beta\gamma}&=&
-\frac{1}{4}g^{\alpha\beta}g^{\gamma\varsigma}g^{\eta\varsigma}
(\mathcal{Q}_{\alpha\eta\varsigma}+\mathcal{Q}_{\eta\varsigma\alpha}
-\mathcal{Q}_{\varsigma\alpha\eta})
\\\nonumber
&\times&(\mathcal{Q}_{\gamma\beta\varsigma}+\mathcal{Q}_{\beta\varsigma\gamma}
-\mathcal{Q}_{\varsigma\gamma\beta}),
\\\nonumber
&=&-\frac{1}{4}(\mathcal{Q}^{\beta\varsigma\gamma}+\mathcal{Q}^{\varsigma\gamma\beta}
-\mathcal{Q}_{\gamma\beta\varsigma})
\\\nonumber
&\times&(\mathcal{Q}_{\gamma\beta\varsigma}+\mathcal{Q}_{\beta\varsigma\gamma}
-\mathcal{Q}_{\varsigma\gamma\beta}),
\\\nonumber
&=&-\frac{1}{4}(2\mathcal{Q}^{\beta\varsigma\gamma}\mathcal{Q}_{\varsigma\gamma\beta}
-
\mathcal{Q}^{\beta\varsigma\gamma}\mathcal{Q}_{\beta\varsigma\gamma}),
\\\nonumber
g^{\alpha\beta}\mathcal{L}^{\gamma}_{\eta\gamma}\mathcal{L}^{\eta}_{\alpha\beta}&=&
\frac{1}{4}g^{\alpha\beta}g^{\eta\varsigma}\mathcal{Q}_{\varsigma}
(\mathcal{Q}_{\beta\alpha\varsigma}+\mathcal{Q}_{\alpha\varsigma\beta}
-\mathcal{Q}_{\varsigma\beta\alpha})
\\\nonumber
&=&\frac{1}{4}\mathcal{Q}^{\varsigma}(2\tilde{\mathcal{Q}_{\varsigma}}-
\mathcal{Q}_{\varsigma}),
\\\nonumber
\mathcal{Q}&=&
-\frac{1}{4}(2\mathcal{Q}^{\beta\varsigma\gamma}\mathcal{Q}_{\gamma\beta\varsigma}
-\mathcal{Q}^{\beta\varsigma\gamma}\mathcal{Q}_{\beta\varsigma\gamma}
-2\mathcal{Q}^{\varsigma}\tilde{\mathcal{Q}_{\varsigma}} +
\mathcal{Q}^{\varsigma}\mathcal{Q}_{\varsigma}).
\end{eqnarray}
According to Eq.(\ref{22}), we obtain
\begin{eqnarray}\nonumber
\mathcal{P}^{\gamma\alpha\beta}&=&\frac{1}{4}[-\mathcal{Q}^{\gamma\alpha\beta}
+
\mathcal{Q}^{\alpha\gamma\beta}+\mathcal{Q}^{\beta\gamma\alpha}+\mathcal{Q}^{\gamma}g^{\alpha\beta}
\\\nonumber
&-&\tilde{\mathcal{Q}^{\gamma}}g^{\alpha\beta}-\frac{1}{2}(g^{\gamma\alpha}
\mathcal{Q}^{\beta}+g^{\gamma\beta}\mathcal{Q}^{\alpha})],
\\\nonumber
-\mathcal{Q}_{\gamma\alpha\beta}\mathcal{P}^{\gamma\alpha\beta} &=&
-\frac{1}{4}[-\mathcal{Q}_{\gamma\alpha\beta}\mathcal{Q}^{\gamma\alpha\beta}
\\\nonumber
&+&\mathcal{Q}_{\gamma\alpha\beta}\mathcal{Q}^{\alpha\gamma\beta}+\mathcal{Q}^{\beta\gamma\alpha}
\mathcal{Q}_{\gamma\alpha\beta}+\mathcal{Q}_{\gamma\alpha\beta}\mathcal{Q}^{\gamma}g^{\alpha\beta}
\\\nonumber
&-&\mathcal{Q}_{\gamma\alpha\beta}\tilde{\mathcal{Q}^{\gamma}}g^{\alpha\beta}
-\frac{1}{2}\mathcal{Q}_{\gamma\alpha\beta}(g^{\gamma\alpha}\mathcal{Q}^{\beta}
+g^{\gamma\beta}\mathcal{Q}^{\alpha})],
\\\nonumber
&=&
-\frac{1}{4}(-\mathcal{Q}_{\gamma\alpha\beta}\mathcal{Q}^{\gamma\alpha\beta}
+2\mathcal{Q}_{\gamma\alpha\beta}\mathcal{Q}^{\alpha\gamma\beta}+\mathcal{Q}^{\gamma}
\mathcal{Q}_{\gamma}-2\tilde{\mathcal{Q}^{\gamma}}\mathcal{Q}_{\gamma}),
\\\nonumber
&=& \mathcal{Q}.
\end{eqnarray}

\section*{Appendix B: Variation of Non-Metricity Scalar}
\renewcommand{\theequation}{B\arabic{equation}}
\setcounter{equation}{0}

All the non-metricity tensors are given as
\begin{eqnarray}\nonumber
\mathcal{Q}_{\gamma\alpha\beta}&=&\nabla_{\gamma}g_{\alpha\beta},
\\\nonumber
\mathcal{Q}^{\gamma}_{\alpha\beta}&=&g^{\gamma\eta}\mathcal{Q}_{\eta\alpha\beta}
=g^{\gamma\eta}\nabla_{\eta}g_{\alpha\beta}
=\nabla^{\gamma}g_{\alpha\beta},
\\\nonumber
\mathcal{Q}_{\gamma
\beta}^{\alpha}&=&g^{\alpha\varsigma}\mathcal{Q}_{\gamma\varsigma\beta}
=g^{\alpha\varsigma}\nabla_{\gamma}g_{\varsigma\beta}
=-g_{\alpha\varsigma}\nabla_{\gamma}g^{\alpha\varsigma},
\\\nonumber
\mathcal{Q}_{\gamma\alpha}^{\beta} &=&
g^{\beta\varsigma}\mathcal{Q}_{\gamma\alpha\varsigma}
=g^{\beta\varsigma}\nabla_{\gamma}g_{\alpha\varsigma}
=-g_{\alpha\varsigma}\nabla_{\gamma}g^{\beta\varsigma},
\\\nonumber
\mathcal{Q}^{\gamma\alpha}_{\beta}&=&
g^{\alpha\varsigma}g^{\gamma\eta}\nabla_{\eta}g_{\varsigma\beta}
=g^{\alpha\varsigma}\nabla^{\gamma}g_{\beta\varsigma}
=-g_{\varsigma\beta}\nabla^{\gamma}g^{\alpha\varsigma},
\\\nonumber
\mathcal{Q}^{\gamma\beta} _{\alpha} &=&
g^{\beta\varsigma}g^{\gamma\eta}\nabla_{\eta}g_{\alpha\varsigma}
=g^{\beta\varsigma}\nabla^{\gamma}g_{\alpha\varsigma}
=-g_{\alpha\varsigma}\nabla^{\gamma}g^{\beta\varsigma},
\\\nonumber
\mathcal{Q}^{\alpha   \beta} _{\gamma} &=&
g^{\alpha\varsigma}g^{\beta\eta}\nabla_{\gamma}g_{\varsigma\eta}
=-g^{\alpha\varsigma}g_{\varsigma\eta}\nabla_{\gamma}g^{\beta\varsigma}
=-\nabla_{\gamma}g^{\alpha\beta}.
\\\nonumber
\mathcal{Q}^{\gamma\alpha\beta}&=&-\nabla^{\gamma}g^{\alpha\beta},
\end{eqnarray}
By using Eq.(\ref{4}), we have
\begin{eqnarray}\nonumber
\delta \mathcal{Q} &=&-\frac{1}{4}
\delta(-\mathcal{Q}^{\gamma\beta\varsigma}
\mathcal{Q}_{\gamma\beta\varsigma}+2\mathcal{Q}^{\gamma\beta\varsigma}
\mathcal{Q}_{\varsigma\gamma\beta}-2\mathcal{Q}^{\varsigma}
\tilde{\mathcal{Q}_{\varsigma}}+\mathcal{Q}^{\varsigma}\mathcal{Q}_{\varsigma}),
\\\nonumber
&=&-\frac{1}{4}(-\delta \mathcal{Q}^{\gamma\beta\varsigma}
\mathcal{Q}_{\gamma\beta\varsigma} -
\mathcal{Q}^{\gamma\beta\varsigma}\delta
\mathcal{Q}_{\gamma\beta\varsigma} + 2\delta
\mathcal{Q}_{\gamma\beta\varsigma}\mathcal{Q}^{\varsigma\gamma\beta}
\\\nonumber
&+& 2 \mathcal{Q}^{\gamma\beta\varsigma}\delta
\mathcal{Q}_{\varsigma\gamma\beta}-2\delta
\mathcal{Q}^{\varsigma}\tilde{\mathcal{Q}_{\varsigma}}+\delta
\mathcal{Q}^{\varsigma}\mathcal{Q}_{\varsigma}-2
\mathcal{Q}^{\varsigma}\delta \tilde {\mathcal{Q}_{\varsigma}} +
\mathcal{Q}^{\varsigma}\delta \mathcal{Q}_{\varsigma}),
\\\nonumber
&=&-\frac{1}{4}[\mathcal{Q}_{\gamma\beta\varsigma}\nabla^{\gamma}\delta
g^{\beta\varsigma}-\mathcal{Q}^{\gamma\beta\varsigma}\nabla_{\gamma}\delta
g_{\beta\varsigma}-2\mathcal{Q}_{\varsigma\gamma\beta}\nabla^{\gamma}\delta
g^{\beta\varsigma}
\\\nonumber
&+&2\mathcal{Q}^{\gamma\beta\varsigma}\nabla_{\varsigma}\delta
g_{\gamma\beta}-2\tilde{\mathcal{Q}_{\varsigma}
\delta}(-g_{\alpha\beta}\nabla^{\varsigma}g^{\alpha\beta})
\\\nonumber
&-&2\mathcal{Q}^{\varsigma}\delta (\nabla^{\eta}g_{\varsigma\eta}) +
\mathcal{Q}_{\varsigma} \delta
(-g_{\alpha\beta}\nabla^{\varsigma}g^{\alpha\beta}) +
\mathcal{Q}^{\varsigma}\delta
(-g_{\alpha\beta}\nabla_{\varsigma}g^{\alpha\beta})],
\\\nonumber
&=&-\frac{1}{4}[\mathcal{Q}_{\gamma\beta\varsigma}\nabla^{\gamma}\delta
g^{\beta\varsigma}-\mathcal{Q}^{\gamma\beta\varsigma}\nabla_{\gamma}\delta
g_{\beta\varsigma}-2\mathcal{Q}_{\varsigma\gamma\beta}\nabla^{\gamma}\delta
g^{\beta\varsigma}
\\\nonumber
&+&2\mathcal{Q}^{\gamma\beta\varsigma}\nabla_{\varsigma}\delta
g_{\gamma\beta}+ 2
\tilde{\mathcal{Q}_{\varsigma}}g^{\alpha\beta}\nabla^{\varsigma}\delta
g_{\alpha\beta}
\\\nonumber
&-&2\mathcal{Q}^{\varsigma}\nabla^{\eta}\delta
g_{\varsigma\eta}+2\tilde{\mathcal{Q}_{\varsigma}}
g_{\alpha\beta}\nabla^{\varsigma}\delta
g^{\alpha\beta}-\mathcal{Q}_{\varsigma}\nabla^{\eta}g^{\alpha\beta}\delta
g_{\alpha\beta}
\\\nonumber
&-&\mathcal{Q}_{\varsigma}g_{\alpha\beta}\nabla^{\varsigma}\delta
g^{\alpha\beta}-\mathcal{Q}_{\varsigma} g^{\alpha\beta}
\nabla_{\varsigma} \delta g_{\alpha\beta}
-\mathcal{Q}^{\varsigma}g_{\alpha\beta}\nabla_{\varsigma}\delta
g_{\alpha\beta}].
\end{eqnarray}
We use the following relations to simplify the above equation
\begin{eqnarray}\nonumber
\delta g_{\alpha\beta}&=&-g_{\alpha\gamma} \delta
g^{\gamma\eta}g_{\eta\beta}-\mathcal{Q}^{\gamma\beta\varsigma}
\nabla_{\gamma}\delta g_{\beta\varsigma},
\\\nonumber
\delta g_{\beta\varsigma}&=&-\mathcal{Q}^{\gamma\beta\varsigma}
\nabla_{\gamma}(-g_{\beta\alpha}\delta
g^{\alpha\eta}g_{\eta\varsigma}),
\\\nonumber
&=&2\mathcal{Q}^{\gamma\beta}_{\varsigma}\mathcal{Q}_{\gamma\beta\alpha}\delta
g^{\alpha\varsigma} +
\mathcal{Q}_{\gamma\eta\varsigma}\nabla^{\gamma}g^{\alpha\varsigma}
\\\nonumber
&=&2\mathcal{Q}^{\gamma\eta}_{\beta}\mathcal{Q}_{\gamma\eta\beta}
\delta g^ {\alpha\beta} +
\mathcal{Q}_{\gamma\beta\varsigma}\nabla^{\gamma}
g^{\beta\varsigma},
\\\nonumber
2\mathcal{Q}^{\gamma\beta\varsigma}\nabla_{\varsigma} \delta
g_{\gamma\beta}&=&
-4\mathcal{Q}_{\alpha}^{\eta\varsigma}\mathcal{Q}_{\varsigma\eta\beta}
\delta g^ {\alpha\beta} _2
\mathcal{Q}_{\beta\varsigma\gamma}\nabla^{\gamma}\delta
g^{\beta\varsigma},
\\\nonumber
-2\mathcal{Q}^{\varsigma} \nabla_{\eta} \delta
g_{\varsigma\eta}&=&2\mathcal{Q}^{\gamma}
\mathcal{Q}_{\beta\gamma\alpha} \delta g^{\alpha\beta}+
2\mathcal{Q}_{\alpha}\tilde{ \mathcal{Q}_{\beta}} \delta
g^{\alpha\beta}
\\\nonumber
&+&2\mathcal{Q}_{\beta}g_{\gamma\varsigma}\nabla^{\gamma}g^{\beta\varsigma}.
\end{eqnarray}
Thus, we have
\begin{equation}\nonumber
\delta
\mathcal{Q}=2\mathcal{P}_{\gamma\beta\varsigma}\nabla^{\gamma}\delta
g^{\beta\varsigma}(\mathcal{P}_{\alpha\gamma\eta}\mathcal{Q}_{\beta}
^{\gamma\eta}-2
\mathcal{P}_{\gamma\eta\beta}\mathcal{Q}^{\gamma\eta}
_{\beta})\delta g^{\alpha\beta},
\end{equation}
where
\begin{eqnarray}\nonumber
2\mathcal{P}{\gamma\beta\varsigma}&=&-\frac{1}{4}[2\mathcal{Q}
_{\gamma\beta\varsigma}-2\mathcal{Q}_{\varsigma\gamma\beta}
-2\mathcal{Q}_{\beta\varsigma\gamma}
\\\nonumber
&+&2(\tilde{\mathcal{Q}_{\gamma}-\mathcal{Q}_{\gamma}})
g_{\beta\varsigma}+2\mathcal{Q}_{\beta}g_{\gamma\eta}],
\\\nonumber
4(\mathcal{P}_{\alpha\gamma\eta}\mathcal{Q}_{\beta} ^{\gamma\eta} -2
\mathcal{P}_{\gamma\eta\beta}\mathcal{Q}^{\gamma\eta}
_{\beta})&=&2\mathcal{Q}^{\gamma\eta}
_{\beta}\mathcal{Q}_{\gamma\eta\alpha}-4\mathcal{Q}_{\alpha}
^{\gamma\eta}\mathcal{Q}_{\eta\gamma\beta}+2\mathcal{Q}_{\gamma\alpha\beta}
\tilde{\mathcal{Q}^{\gamma}}
\\\nonumber
&-&\mathcal{Q}^{\gamma}\mathcal{Q}_{\gamma\alpha\beta}
+2\mathcal{Q}^{\gamma}\mathcal{Q}_{\beta\gamma\alpha}+2\mathcal{Q}
_{\alpha}\tilde{\mathcal{Q}_{\beta}},
\end{eqnarray}
\\
\textbf{Data availability:} This research did not generate or
analyze any new data.

\end{document}